\documentclass[12pt]{iopart}
\usepackage{mathrsfs}
\usepackage{enumerate}
\usepackage{bbm}
\expandafter\let\csname equation*\endcsname\relax

\expandafter\let\csname endequation*\endcsname\relax
\usepackage{amsmath}
\usepackage{amsthm}
\usepackage{amsbsy}
\usepackage{amsfonts}
\usepackage{physics}
\usepackage{hyperref}
\usepackage{cleveref}
\usepackage{ulem} 
\usepackage{xcolor}
\numberwithin{equation}{section}
\renewcommand{\tr}{\operatorname{Tr}}
\newcommand{\Del}{\nabla}
\newcommand{\del}{\partial}
\renewcommand{\tilde}{\widetilde}
\newcommand{\pos}{\operatorname{Pos}}
\newcommand{\herm}{\operatorname{Herm}}
\newcommand{\psd}{\operatorname{PSD}}
\newcommand{\defi}{:=}

\newcommand{\reals}{\mathbb{R}}

\renewcommand{\d}{\dd\hspace{-0.04em}}

\usepackage{bbm}
\newcommand{\Id}{\operatorname{id}}
\newcommand{\id}{\mathbbm{1}}

\newtheorem{theorem}{Theorem}[section]

\theoremstyle{remark}
\newtheorem{remark}[theorem]{Remark}

\begin{document}

\title{Geometric conditions for saturating the data processing inequality}
\author{Samuel S. Cree and Jonathan Sorce}
\address{Stanford Institute for Theoretical Physics}
\ead{\mailto{scree@stanford.edu}, \mailto{jsorce@stanford.edu}}
\begin{abstract}
	The data processing inequality (DPI) is a scalar inequality satisfied by distinguishability measures on density matrices.
	For some distinguishability measures, saturation of the \emph{scalar} DPI implies an \emph{operator} equation relating the arguments of the measure.
	These results are typically derived using functional analytic techniques.
	In a complementary approach, we use geometric techniques to derive a formula that gives an operator equation from DPI saturation for \emph{any} distinguishability measure; moreover, for a broad class of distinguishability measures, the derived operator equation is sufficient to imply saturation as well. 
	Our operator equation coincides with known results for the sandwiched R\'{e}nyi relative entropies, and gives new results for $\alpha$-$z$ R\'{e}nyi relative entropies and a family of of quantum $f$-divergences, which we compute explicitly.
\end{abstract}

\maketitle

\section{Introduction}

The distinguishability of two pure quantum states $\ket{\psi}$ and $\ket{\phi}$ is completely characterized by the inner product $\braket{\psi}{\phi}.$ For mixed states $\rho$ and $\sigma$, however, characterizing the ``distinguishability'' of the states is more complicated. As a result, there is a zoo of distinguishability measures (see e.g. \cite{uhlmann_transition_1976,uhlmann_transition_1985,wilde_strong_2014,muller-lennert_quantum_2013,matsumoto_new_2018,matsumoto_reverse_2010,cree_fidelity_2020,kholevo_quasiequivalence_1972,hiai_different_2017,hiai_quantum_2011,gao_recoverability_2020,umegaki_conditional_1962,audenaert_alpha-z-relative_2015,petz_quasi-entropies_1985}), each with their own advantages and disadvantages. To qualify as a distinguishability measure, a function of two density matrices really only needs to satisfy one physical principle: it should not be possible for two states to become \emph{more} distinguishable by applying the same quantum channel to both states. This condition is known as the \emph{data processing inequality}, or \emph{DPI}.

Formally, a distinguishability measure $\mathcal{B}$ for states on a Hilbert space $\mathcal{H}$ is a function\footnote{Here $\pos(\mathcal{H})$ denotes the set of positive definite operators, i.e., the set of operators $P$ such that $\ev{P}{\psi} > 0$ holds for all nonzero $\ket{\psi} \in \mathcal{H}.$}
\begin{align}
	\mathcal{B}: \pos(\mathcal{H}) \times \pos(\mathcal{H}) \to \mathbb{R}
\end{align}
that satisfies the data processing inequality: for any quantum channel\footnote{Recall that a quantum channel is a completely positive, trace-preserving, linear map on the space of operators.} $\Lambda $, and any two (generally non-normalized) states $\rho, \sigma  \in \pos(\mathcal{H})$, we have
\begin{align}
	\mathcal{B}(\rho, \sigma ) \geq \mathcal{B}(\Lambda (\rho ),\Lambda (\sigma )).
	\label{eq:DPI}
\end{align}
We will also assume, as part of our definition, that $\mathcal{B}$ is differentiable; this is the case for every distinguishability measure we consider in this paper.

We say that $\Lambda$ is \emph{recoverable} on the states $\rho$ and $\sigma$ if there exists a channel $\mathcal{R}$ satisfying
\begin{equation}
	[\mathcal{R} \circ \Lambda](\rho) = \rho
\end{equation}
and
\begin{equation}
	[\mathcal{R} \circ \Lambda](\sigma) = \sigma.
\end{equation}
(Such a map is sometimes also called ``reversible'' or ``sufficient'' on the pair $\rho, \sigma$.)
When such a channel exists, applying the DPI with respect to $\Lambda$ and then with respect to $\mathcal{R}$ shows that we must have equality in \eqref{eq:DPI} for any distinguishability measure $\mathcal{B}.$
For certain choices of distinguishability measure, the converse is also true; if the distinguishability of $\rho$ and $\sigma$ does not change under the application of $\Lambda$, then $\Lambda$ is recoverable on those states. This fact, that saturation of DPI is equivalent to recoverability for certain distinguishability measures, is known as \emph{Petz recovery} in honor of Petz's proof that the relative entropy is one distinguishability measure with this property \cite{petz1}.

Part of the difficulty in proving Petz recovery for a given distinguishability measure is that saturation of the DPI is a scalar equation, while the result to be proved is an equality of operators -- the initial state $\rho$ must equal the final state $[\mathcal{R} \circ \Lambda](\rho)$.
For this reason, several authors have worked to derive operator equations that are implied by (or equivalent to) saturation of the data-processing inequality \cite{petz_quasi-entropies_1985,jencova_renyi_2018,jencova_renyi_2018-1,jencova_unified_2010,jencova_preservation_2017,jia_petz_2020,wang_revisiting_2020,hiai_different_2017,hiai_quantum_2011,matsumoto_new_2018,gao_recoverability_2020,chehade_saturating_2020,zhang_equality_2020,leditzky_data_2016}.
The techniques for deriving these equations generally fall under the mathematical umbrella of functional analysis; the purpose of this paper is to introduce a complementary, geometric toolkit for deriving operator equations from DPI saturation. 
As we will see, this approach immediately reproduces known results for the relative entropy and fidelity, and more generally for the full class of sandwiched R\'enyi relative entropies.
It also allows us to derive new operator equalities implied by DPI saturation for any smooth distinguishability measure; we compute these identities explicitly for the $\alpha$-$z$ R\'{e}nyi relative entropies and a general family of $f$-divergences.

The basic idea is as follows.
Because a distinguishability measure $\mathcal{B}$ has as its domain two operator manifolds, its \emph{gradient} with respect to either argument is a tangent vector on the corresponding operator manifold.
Tangent vectors on manifolds of operators are themselves operators.
For density matrices $\rho, \sigma$ that saturate DPI for a particular quantum channel $\Lambda$, the function
\begin{equation} \label{eq:positive-f}
	f_{\Lambda}(\rho, \sigma) = \mathcal{B}(\rho, \sigma) - \mathcal{B}(\Lambda(\rho), \Lambda(\sigma))
\end{equation}
is at an extreme value -- its minimum, zero.
As such, the gradient of $f$ with respect to either of its arguments must vanish.
Since the gradient is an operator, this implies an operator equation.
The technical matter of this paper is primarily in (i) explicitly computing this vanishing-gradient equation for specific examples, and (ii) showing that for a broad class of distinguishability measures, this vanishing-gradient equation is also sufficient to imply DPI saturation.

The plan of the paper is as follows.

In section \ref{sec:matrix-manifolds}, we give a refresher on how to take derivatives of functions defined on manifolds of operators.
We give several explicit formulas, the most nontrivial of which is derived in \ref{app:locally-analytic},  that are then used in section \ref{sec:main}.
Readers who are already familiar with derivatives on matrix manifolds may safely skip section \ref{sec:matrix-manifolds} and refer to it only as needed while reading section \ref{sec:main}.

In section \ref{sec:main}, we detail the ``vanishing gradient'' argument alluded to above.
We give the vanishing-gradient equation a simple form in theorem \ref{thm:main-theorem}, and provide a condition under which the vanishing-gradient equation is \emph{equivalent} to DPI saturation, rather than merely implied by it.
We then use theorem \ref{thm:main-theorem} to derive necessary and sufficient conditions for various distinguishability measures to saturate the data processing inequality.
In particular, we replicate a previously known result for the sandwiched R\'enyi relative entropies \cite{leditzky_data_2016}, though our proof technique is different and, for those less familiar with functional analytic techniques, hopefully more intuitive.
We also comment on the case of the $\alpha$-$z$ R\'{e}nyi relative entropies, where the vanishing-gradient equation is not superficially identical to either of the DPI saturation conditions derived in \cite{zhang_equality_2020,chehade_saturating_2020}.

In section \ref{sec:boundary}, we derive operator equations implied by DPI saturation in the case that one or both of the density matrices are positive semidefinite but not strictly positive.
The basic idea is that while a general distinguishability measure will not be differentiable in a neighborhood of a density matrix with a vanishing eigenvalue, nor will it necessarily even satisfy DPI in a neighborhood of that density matrix, directional derivatives along the boundary of the space of positive operators are required to vanish.

In section \ref{sec:discussion}, we comment on potential applications to Petz recovery and other directions for future work.

\section{User's guide to derivatives on matrix manifolds} \label{sec:matrix-manifolds}

Let $\mathcal{H}$ be an $n$-dimensional Hilbert space.
The set of Hermitian operators on $\mathcal{H}$, denoted $\herm(\mathcal{H})$, is an $n^2$-dimensional real manifold with coordinates given by the real and imaginary parts of the independent matrix entries under any choice of basis for $\mathcal{H}$.
The space of positive operators, $\pos(\mathcal{H})$, is an $n^2$-dimensional real submanifold of $\herm(\mathcal{H})$. (Because the eigenvalues of a Hermitian matrix are continuous functions of the matrix entries, the eigenvalues of a strictly positive matrix remain strictly positive under a small Hermitian perturbation; so $\pos(\mathcal{H})$ is an open subset of $\herm(\mathcal{H})$, and hence a submanifold.) Our goal in this section will be to understand the derivatives of functions $f : \pos(\mathcal{H}) \rightarrow \reals$.
We develop this machinery so that for any distinguishability measure $\mathcal{B} : \pos(\mathcal{H}) \times \pos(\mathcal{H}) \rightarrow \reals$, and any positive operator $\sigma$, we will be able to compute the derivative of the restricted map $\mathcal{B}|_{\sigma} : \pos(\mathcal{H}) \rightarrow \reals$ defined by $\mathcal{B}|_{\sigma}(\rho) = \mathcal{B}(\rho, \sigma).$ Our pedagogy in this section roughly follows section 5 of \cite{CGW}.

Let $f : M \rightarrow N$ be a smooth map between manifolds $M$ and $N$.
The derivative of $f$ at a point $p \in M$, denoted $\d f|_{p}$, is defined as a linear map from the tangent space $T_p M$ to the tangent space $T_{f(p)} N.$ Its action on the tangent space is such that for any curve $\gamma$ in $M$ passing through $p$, the derivative $\d f|_{p}$ maps the tangent vector of $\gamma$ at the point $p \in M$ to the tangent vector of $f(\gamma)$ at the point $f(p) \in N$.

Now, we restrict to the case where $f$ is a map from $\pos(\mathcal{H})$ to the real numbers.
The tangent space to $\pos(\mathcal{H})$ at any point $\rho \in \pos(\mathcal{H})$ is isomorphic to the set of Hermitian operators $\herm(\mathcal{H}).$\footnote{A tangent vector to $\rho$ in $\pos(\mathcal{H})$ is a matrix $M$ such that $\rho + \epsilon M$ is still in $\pos(\mathcal{H})$ for sufficiently small $\epsilon$.
This is true if and only if the eigenvalues of $M$ are real, i.e., if and only if $M$ is Hermitian.} The tangent space to $\reals$ at any point $f(\rho) \in \reals$ is isomorphic to the real numbers $\reals$.
So $\d f|_{\rho} : T_\rho \pos(\mathcal{H}) \rightarrow T_{f(\rho)} \reals$ can be thought of as a linear map $\d f|_{\rho} : \herm(\mathcal{H}) \rightarrow \reals.$ But using the Hilbert-Schmidt inner product, any linear map from $\herm(\mathcal{H})$ to $\reals$ can be written as a Hermitian operator! More specifically, there will always be a unique Hermitian operator $\Del f|_{\rho} \in \herm(\mathcal{H})$ satisfying\footnote{For the more geometrically inclined: the derivative of $f$ at $\rho$ is in the dual space $T^*_{\rho}\pos(\mathcal{H}).$ The gradient $\Del f$ is obtained by ``raising the index'' by mapping $T^*_{\rho} \pos(\mathcal{H}) \rightarrow T_{\rho} \pos(\mathcal{H})$ isomorphically using the Hilbert-Schmidt inner product on $T_{\rho} \pos(\mathcal{H})$.}
\begin{equation} \label{eq:dualize}
	\langle\Del f|_{\rho},  M\rangle_{\text{HS}} = \d f|_{\rho}(M)
\end{equation}
for all $M \in \herm(\mathcal{H}).$ Here the Hilbert-Schmidt inner product on Hermitian matrices $A, B$ is defined by
\begin{equation}
	\langle A, B \rangle_{\text{HS}} = \tr(A B).
\end{equation}

As explained in the introduction and detailed in section \ref{sec:main}, saturation of the data processing inequality for a distinguishability measure $\mathcal{B}$ implies certain operator equations for derivatives of related functions $f : \pos(\mathcal{H}) \rightarrow \reals$. To put these equations in a useful analytic form, we will need to be able to compute the Hermitian operator $\Del f|_{\rho}$ appearing in equation \eqref{eq:dualize}. This is done by computing $\d f|_{\rho}$, then dualizing with respect to the Hilbert-Schmidt inner product.
To compute $\d f$, it is helpful to write $f$ as a composition of simple functions $f = f_1 \circ \dots \circ f_n$, then to compute the derivative of $f$ using the chain rule.

Every distinguishability measure we study in section \ref{sec:main} can be decomposed into pieces that are either \emph{linear} or \emph{locally analytic}. Denoting by $\mathcal{L}(\mathcal{H})$ the full space of linear operators on $\mathcal{H}$, any linear map $f_{j}$ between vector subspaces\footnote{Note that since the $f_j$ are intermediary functions that must \emph{compose} to some function $f : \pos(\mathcal{H}) \rightarrow \reals$, we can allow them to have arbitrary domains and codomains within $\mathcal{L}(\mathcal{H})$. In particular, left-multiplication by a fixed operator is a linear map from $\mathcal{L}(\mathcal{H})$ to itself, and the trace is a linear map from $\mathcal{L}(\mathcal{H})$ to $\reals.$} of $\mathcal{L}(\mathcal{H})$ has derivative given by
\begin{equation} \label{eq:linear-derivative}
	\d f_{j}|_{A}(M) = f_j(M),
\end{equation}
which is straightforward to derive from the formula
\begin{equation} \label{eq:derivative-definition}
	\d f_{j}|_{A}(M) = \lim_{\epsilon \rightarrow 0} \frac{f_j(A + \epsilon M) - f_j(A)}{\epsilon}.
\end{equation}
Note that the derivative is independent of the point $A \in \mathcal{L}(\mathcal{H})$ where the derivative is taken; the fact that the derivative of a real linear function $y(x) = a x + b$ is independent of $x$ is a special case of this more general principle.

We call a function $f_j$ from $\herm(\mathcal{H})$ to itself \emph{locally analytic} at $A \in \herm(\mathcal{H})$ if, in a neighborhood of $A$, $f$ can be written as a Taylor series centered at a multiple of the identity:
\begin{equation}
	f(A) = \sum_{m=0}^{\infty} c_m (A - \alpha I)^{m}.
\end{equation}
Examples include $f(A) = e^{A}$ for $A$ Hermitian, and $f(A) = A^{\alpha}$ or $f(A) = \log(A)$ for $A$ positive and $\alpha$ real. The derivative of a locally analytic function is given by the formula
\begin{equation} \label{eq:locally-analytic-formula}
	\d f|_{A}(M)
		= \sum_{j} f'(\lambda_{j}) \Pi_{j} M \Pi_{j}
			+ \sum_{j \neq k} \frac{f(\lambda_{j}) - f(\lambda_{k})}{\lambda_j - \lambda_k} \Pi_{j} M \Pi_{k},
\end{equation}
where $A = \sum_{j} \lambda_{j} \Pi_j$ is the spectral decomposition of $A$. We give a derivation of this formula in \ref{app:locally-analytic}.

Finally, we note that derivatives of matrix functions satisfy sum, product, and chain rules analogous with those of single-variable calculus. The sum and product rules are given by
\begin{equation} \label{eq:sum-rule}
	\d (f + g)|_{A}(M)
		= \d f|_{A}(M) + \d g|_{A}(M)
\end{equation}
and
\begin{equation} \label{eq:product-rule}
	\d (f \cdot g)|_{A}(M)
		= \d f|_{A}(M) g(A)+ f(A) \d g|_{A}(M).
\end{equation}
The chain rule says that if $f_2 : Q \rightarrow R$ and $f_1 : R \rightarrow S$ are maps of matrix manifolds, then for a point $q \in Q$ and a tangent vector $v \in T_q Q$, the composition $(f_1 \circ f_2)$ satisfies
\begin{equation} \label{eq:chain-rule}
	\d (f_1 \circ f_2)|_{q}(v)
		= \d f_1|_{f_2(q)}[\d f_2|_{q}(v)].
\end{equation}
All three of these rules can be derived from equation \eqref{eq:derivative-definition} using the same proof techniques as are used to derive their analogues in single-variable calculus.

Note that even when $f_1 \circ f_2$ is a map from $\pos(H)$ to $\reals$, there is not a general chain rule for the matrix gradient $\Del (f_1 \circ f_2)|_{\rho}$ (cf.
equation \eqref{eq:dualize}).
This is because the gradient is only defined for matrix functions whose codomain is $\reals$; even though $f_1 \circ f_2$ and $f_1$ both have codomain $\reals$, the codomain of $f_2$ will generally be a matrix manifold.
When we compute the matrix gradients of quantum distinguishability measures in section \ref{sec:main}, we will always compute $\d f$ directly using the chain rule, then dualize to find $\Del f.$

\section{Saturation of DPI for positive definite matrices} \label{sec:main}

In this section, we show that several known ``DPI saturation $\Leftrightarrow$ operator equation'' results for particular distinguishability measures $\mathcal{B}$ are special cases of the universal equation
\begin{equation} \label{eq:universal-derivative-equation}
	\d_{1}(\mathcal{B} - \mathcal{B} \circ (\Lambda \times \Lambda))|_{\rho, \sigma} = 0,
\end{equation}
where the triple $(\rho, \sigma, \Lambda)$ saturates the data processing inequality.
We also give some examples of new conditions for equality arising from this equation.

In the first subsection, we explain this equation; we also put it in a more explicit form in theorem \ref{thm:main-theorem}.
In the second subsection, we apply theorem \ref{thm:main-theorem} to several specific distinguishability measures.
We reproduce known results for the sandwiched R\'{e}nyi relative entropies, and new results for a family of quantum $f$-divergences.
We also calculate equation \eqref{eq:universal-derivative-equation} for the general $\alpha$-$z$ R\'{e}nyi relative entropies studied in \cite{audenaert_alpha-z-relative_2015}, and compare with previous results from \cite{chehade_saturating_2020,zhang_equality_2020}.

\subsection{Main theorem}

As defined in the introduction, a distinguishability measure $\mathcal{B}$ on quantum states is a map $\mathcal{B} : \pos(\mathcal{H}) \times \pos(\mathcal{H}) \rightarrow \reals$ such that (i) $\mathcal{B}$ is differentiable as a matrix function in either of its arguments, and (ii) $\mathcal{B}$ satisfies the data processing inequality
\begin{equation}
	\mathcal{B}(\rho, \sigma) \geq \mathcal{B}(\Lambda(\rho), \Lambda(\sigma))
\end{equation}
for any quantum channel $\Lambda$. 

Let us define a new function $f_{\Lambda} : \pos(\mathcal{H}) \times \pos(\mathcal{H}) \rightarrow \reals$ given by
\begin{equation} \label{eq:special-f}
	f_{\Lambda}(\rho, \sigma) = \mathcal{B}(\rho, \sigma) - \mathcal{B}(\Lambda(\rho), \Lambda(\sigma)).
\end{equation}
The function $f_{\Lambda}$ is differentiable, since both of its terms are differentiable, and is bounded below by zero.
So when $f_{\Lambda}$ reaches its minimum -- i.e., when the data processing inequality is saturated -- we must have $\d_{(1)} f_{\Lambda}|_{\rho, \sigma} = \d_{(2)} f_{\Lambda}|_{\rho, \sigma} = 0$, where $\d_{(1)}$ and $\d_{(2)}$ signify the manifold derivatives of $f$ with respect to its first or second argument, respectively.
As explained in section \ref{sec:matrix-manifolds}, in the discussion surrounding equation \eqref{eq:dualize}, vanishing of $\d_{(j)} f$ is equivalent to the vanishing of its Hilbert-Schmidt dual $\Del_{(j)} f$. So when $(\rho, \sigma, \Lambda)$ saturates the data processing inequality for the distinguishability measure $\mathcal{B}$, the operators $\Del_{(j)} f$ must be identically zero.

In the following theorem, we provide a more convenient formula for the equation $\Del_{(j)} f_{\Lambda}|_{\rho, \sigma} = 0$ and give a condition for when this equation is equivalent to DPI saturation, rather than only \emph{implied by} DPI saturation.
We prove this theorem in a very general setting, with the function $\mathcal{B}$ only required to satisfy DPI and differentiability in a neighborhood of two fixed Hermitian operators. However, our primary application is to distinguishability measures, which are smooth and satisfy the DPI for any positive operators.

\begin{theorem} \label{thm:main-theorem}
Let $\mathcal{H}$ be a finite-dimensional Hilbert space, and $\Lambda : \herm(\mathcal{H}) \rightarrow \herm(\mathcal{H})$ a linear map on Hermitian operators.
(In particular, $\Lambda$ can be a quantum channel.) Suppose that $\mathcal{B}$ is a map from $\herm(\mathcal{H}) \times \herm(\mathcal{H})$ to $\reals$, and $\rho, \sigma$ are operators in $\herm(\mathcal{H})$ such that $\mathcal{B}$ satisfies the data processing inequality with respect to $\Lambda$ in a neighborhood of $(\rho, \sigma)$. I.e., for any Hermitian operators $M_1, M_2$ and $\epsilon$ sufficiently small, $\mathcal{B}$ must satisfy
\begin{equation} \label{eq:DPI-neighborhood}
	\mathcal{B}(\rho + \epsilon M_1, \sigma + \epsilon M_2) \geq \mathcal{B}(\Lambda(\rho + \epsilon M_1), \Lambda(\sigma + \epsilon M_2)).
\end{equation}
Furthermore, suppose that $\mathcal{B}$ is differentiable with respect to either of its arguments in a neighborhood of $(\rho, \sigma),$ and that the data processing inequality is saturated at $(\rho, \sigma)$:
\begin{equation}
	\mathcal{B}(\rho, \sigma) = \mathcal{B}(\Lambda(\rho), \Lambda(\sigma)).
\end{equation}
Then:
\begin{enumerate}
	\item
	The operator equations
	\begin{align}
		\Del_{(1)} \mathcal{B}|_{\rho, \sigma}
			& = \Lambda^* \left( \Del_{(1)} \mathcal{B}|_{\Lambda(\rho), \Lambda(\sigma)} \right), \label{eq:first-vanishing-gradient} \\
		\Del_{(2)} \mathcal{B}|_{\rho, \sigma}
			& = \Lambda^* \left( \Del_{(2)} \mathcal{B}|_{\Lambda(\rho), \Lambda(\sigma)} \right) \label{eq:second-vanishing-gradient}
	\end{align}
	are satisfied, where $\Del_{(j)}$ denotes the matrix gradient with respect to the $j$-th argument of $\mathcal{B}$, and $\Lambda^*$ is the adjoint of $\Lambda$ with respect to the Hilbert-Schmidt inner product.

	(If $\mathcal{B}$ is only differentiable with respect to its $j$-th argument, then the corresponding equation still holds.)

	\item
	If $\d_{(1)} \mathcal{B}$ satisfies
	\begin{equation} \label{eq:converse-condition-1}
		\d_{(1)} \mathcal{B}|_{\rho, \sigma}(\rho)
		= g \left[ \mathcal{B}(\rho, \sigma) \right],
	\end{equation}
	and
	\begin{equation} \label{eq:converse-condition-2}
		\d_{(1)} \mathcal{B}|_{\Lambda(\rho), \Lambda(\sigma)}[\Lambda(\rho)]
		= g \left( \mathcal{B}[\Lambda(\rho), \Lambda(\sigma)]  \right),
	\end{equation}
	for some invertible function $g$, then equation \eqref{eq:first-vanishing-gradient} implies that $\mathcal{B}$ saturates the data processing inequality with respect to $\Lambda$ at the point $(\rho, \sigma).$ An analogous condition holds for $\d_{(2)} \mathcal{B}$ and equation \eqref{eq:second-vanishing-gradient}.
\end{enumerate}
\end{theorem}
\begin{remark}
	The conditions \eqref{eq:converse-condition-1} and \eqref{eq:converse-condition-2} may seem odd, but they are quite natural features for a distinguishability measure. 
	The derivative $d_{(1)} \mathcal{B}|_{\rho, \sigma}$ evaluated on $\rho $ measures how the value of the distinguishability measure would change under a perturbation of the form $\rho  \to ( 1+ \varepsilon )\rho $.
	Many functions have some well-defined transformation under scalar multiplication, such as $\mathcal{B}(k \rho ,\sigma ) = a(k ) \mathcal{B}( \rho ,\sigma ) + b(k )$ for a positive constant $k $ and some functions $a(k )$ and $b(k )$.
	If $a$ is nonzero, then this is an affine, invertible function of $\mathcal{B}( \rho ,\sigma ) $.
	More generally, these conditions just mean that the distinguishability changes in some straightforward, predictible manner when one of its arguments is scaled by a constant.
\end{remark}

\begin{proof}
\begin{enumerate}
	\item
	We have already explained in the preamble to this subsection why the equation $\d_{(1)} f_{\Lambda}(\rho, \sigma) = \d_{(2)} f_{\Lambda}(\rho, \sigma) = 0$ is implied by saturation of DPI for $f_{\Lambda}$ defined as in equation \eqref{eq:special-f}.
	We need only show that this is equivalent to equations \eqref{eq:first-vanishing-gradient}, \eqref{eq:second-vanishing-gradient}.

	Without loss of generality, we will restrict our attention to the first-argument derivative $d_{(1)}$, and for simplicity of notation we will also temporarily drop the $(1)$ subscript.
	We rewrite $f_{\Lambda}$ from equation \eqref{eq:special-f} as
	\begin{equation}
		f_{\Lambda}|_{\rho, \sigma} = \mathcal{B}|_{\rho, \sigma} - \left[\mathcal{B} \circ (\Lambda \times \text{id})\right]|_{\rho, \Lambda(\sigma)}.
	\end{equation}
	The action of $\d f_{\Lambda }$ on a Hermitian operator $M$ can be written using the chain rule (equation \eqref{eq:chain-rule}) as
	\begin{equation}
		\d f_{\Lambda}|_{\rho, \sigma}(M)
			= \d \mathcal{B}|_{\rho, \sigma}(M) - \d \mathcal{B}|_{\Lambda(\rho), \Lambda(\sigma)} \left[d (\Lambda \times \Id)|_{\rho, \Lambda(\sigma)}(M)\right].
	\end{equation}
	The map $\Lambda \times \Id$ is linear, so thanks to equation \eqref{eq:linear-derivative} we know that its derivative satisfies $\d (\Lambda \times \Id) = \Lambda.$ So we have
	\begin{equation} \label{eq:theorem-penultimate}
		\d f_{\Lambda}|_{\rho, \sigma}(M)
			= \d \mathcal{B}|_{\rho, \sigma}(M) - \d \mathcal{B}|_{\Lambda(\rho), \Lambda(\sigma)} \left[\Lambda(M)\right].
	\end{equation}

	To define the gradient $\Del$, we dualize with respect to the Hilbert-Schmidt inner product using equation \eqref{eq:dualize}.
	The gradient $\Del \mathcal{B}|_{\rho, \sigma}$ is the unique Hermitian operator satisfying
	\begin{equation}
		\tr(\Del \mathcal{B}|_{\rho, \sigma} M) = \d \mathcal{B}|_{\rho, \sigma}(M)
	\end{equation}
	for all $M$.
	Using this definition, we rewrite equation \eqref{eq:theorem-penultimate} as
	\begin{equation} \label{eq:theorem-penultimate2}
		\d f_{\Lambda}|_{\rho, \sigma}(M)
			= \tr(\Del \mathcal{B}|_{\rho, \sigma} M) - \tr[\Del \mathcal{B}|_{\Lambda(\rho), \Lambda(\sigma)} \Lambda(M)].
	\end{equation}
	In terms of the adjoint of $\Lambda$, defined by $\tr[\Lambda^*(A) B] = \tr[A \Lambda(B)]$ for all Hermitian $A$ and $B$, we have
	\begin{equation}
		\d f_{\Lambda}|_{\rho, \sigma}(M)
			= \tr(\Del \mathcal{B}|_{\rho, \sigma} M) - \tr[\Lambda^*\left(\Del \mathcal{B}|_{\Lambda(\rho), \Lambda(\sigma)}\right) M].
	\end{equation}
	Since saturation of the DPI implies that this equation vanishes for all $M$, we have (reinserting the subscript $(1)$ to denote the derivative with respect to the first argument)
	\begin{equation}
		\Del_{(1)} \mathcal{B}|_{\rho, \sigma} = \Lambda^*\left(\Del_{(1)} \mathcal{B}|_{\Lambda(\rho), \Lambda(\sigma)}\right),
	\end{equation}
	as desired.

	The same argument holds for the gradient with respect to the second argument of $\mathcal{B}$, so long as $\mathcal{B}$ is differentiable in that argument in a neighborhood of $\rho, \sigma.$

	\item
	In the previous part of this proof, we showed that equation \eqref{eq:first-vanishing-gradient} is equivalent to
	\begin{equation}
		\d_{(1)} \mathcal{B}|_{\rho, \sigma}(M)
			= \d_{(1)} \mathcal{B}|_{\Lambda(\rho), \Lambda(\sigma)} \left[\Lambda(M)\right].
	\end{equation}
	If we set $M = \rho$, this gives
	\begin{equation}
		\d_{(1)} \mathcal{B}|_{\rho, \sigma}(\rho)
			= \d_{(1)} \mathcal{B}|_{\Lambda(\rho), \Lambda(\sigma)} \left[\Lambda(\rho)\right].
	\end{equation}
	Applying equations \eqref{eq:converse-condition-1} and \eqref{eq:converse-condition-2} yields
	\begin{equation}
		g\left[\mathcal{B}(\rho, \sigma) \right]
		= g\left[  \mathcal{B}\left[\Lambda(\rho), \Lambda(\sigma)\right] \right],
	\end{equation}
	which gives $\mathcal{B}(\rho, \sigma) = \mathcal{B}\left[\Lambda(\rho), \Lambda(\sigma)\right]$ by the invertibility of $g$, as desired.

	An analogous argument holds for derivatives with respect to the second argument of $\mathcal{B}.$
\end{enumerate}
\end{proof}

\subsection{Examples} \label{subsec:lightning-proofs}

We now compute equations \eqref{eq:first-vanishing-gradient} and \eqref{eq:second-vanishing-gradient} for particular distinguishability measures to derive the corresponding DPI saturation conditions.
We derive our first two DPI saturation conditions (the relative entropy and fidelity) in some detail to help the reader get a feel for calculating matrix derivatives of distinguishability measures. We then give slightly less detailed computations of the saturation conditions for the full family of sandwiched R\'{e}nyi relative entropies, for the $\alpha$-$z$ R\'{e}nyi relative entropies, and for the quantum $f$-divergences.

\subsubsection{Relative entropy} \label{subsubsec:relative-entropy}

The relative entropy, defined by $D(\rho || \sigma) = \tr(\rho \log \rho) - \tr(\rho \log \sigma),$ satisfies the data processing inequality so long as $\rho$ is positive semidefinite and $\sigma$ is positive.
The case when $\rho$ is not strictly positive will be dealt with in section \ref{sec:boundary}; for the moment, we assume both $\rho$ and $\sigma$ are in $\pos(\mathcal{H}).$

As a function of its first argument, $D$ can be written
\begin{equation}
	D|_{\rho, \sigma}
		= \left[\tr \circ \left(\Id \cdot \log - R_{\log(\sigma)}\right)\right] (\rho),
\end{equation}
where $\Id$ is the identity superoperator and $R_{\log(\sigma)}$ denotes right multiplication by $\log(\sigma).$ By successively applying the chain rule \eqref{eq:chain-rule}, the product rule \eqref{eq:product-rule}, the linearity of the derivative \eqref{eq:sum-rule}, the formula for the derivative of a linear function \eqref{eq:linear-derivative}, and the derivative of the matrix logarithm via equation \eqref{eq:locally-analytic-formula}, we may compute the $\d_{(1)}$ derivative of $D$ as
\begin{equation}
	\d_{(1)} D|_{\rho, \sigma}(M)
		= \tr\left[ \left(\log(\rho) - \log(\sigma) + \id\right) M\right].
\end{equation}
So the gradient of $D$ with respect to its first argument is
\begin{equation} \label{eq:rel-ent-grad}
	\Del_{(1)} D|_{\rho, \sigma} = \log(\rho) - \log(\sigma) + \id.
\end{equation}
By theorem \ref{thm:main-theorem}, DPI saturation implies the operator equation
\begin{equation} \label{eq:relative-entropy-condition}
	\log(\rho) - \log(\sigma)
	= \Lambda^* \left[ \log[\Lambda(\rho)] - \log[\Lambda(\sigma)]\right],
\end{equation}
where we have also used that when $\Lambda$ is a quantum channel, $\Lambda^*$ is unital. This equality was obtained by Petz using operator algebraic methods as theorem 3.1 of \cite{petz2003monotonicity}.

To see that this equation is \emph{equivalent} to DPI saturation, we can simply left-multiply by $\rho$ and take the trace.
The fact that left-multiplying by $\rho$ and taking the trace reproduces the relative entropy is a special case of condition \eqref{eq:converse-condition-1} in theorem \ref{thm:main-theorem}.

For completeness, we also give the vanishing-gradient equation \eqref{eq:second-vanishing-gradient} for the $\d_{(2)}$ derivative of $D$.
As a function of its second argument, $D$ can be written
\begin{equation}
	D|_{\rho, \sigma}
		= \left[[-\Id+\tr(\rho \log \rho)] \circ \tr \circ L_{\rho} \circ \log \right](\sigma),
\end{equation}
where $L_{\rho}$ denotes left-multiplication by $\rho$.
It is straightforward to compute the $\d_{(2)}$ derivative of $D$ using the derivative rules from section \ref{sec:matrix-manifolds}.
The final result of this calculation gives the $\Del_{(2)}$ gradient of $D$ as
\begin{equation}
	\Del_{(2)} D|_{\rho, \sigma}
		= - \sum_{j=k} \frac{1}{p_{j}} \Pi_{j} \rho \Pi_{j} - \sum_{j\neq k} \frac{\log(p_j) - \log(p_k)}{p_j - p_k} \Pi_{k} \rho \Pi_{j},
\end{equation}
where $\sum_{j} p_{j} \Pi_{j}$ is the spectral decomposition of $\sigma.$ The corresponding operator equation \eqref{eq:second-vanishing-gradient} must vanish when DPI is saturated, but this equation is much less elegant than the $\Del_{(1)}$-gradient equation \eqref{eq:relative-entropy-condition}.
Furthermore, it is not obviously \emph{equivalent} to DPI saturation, since the $\d_{(2)}$ derivative of $D$ does not satisfy condition \eqref{eq:converse-condition-2} of theorem \ref{thm:main-theorem}.

\subsubsection{Fidelity}
The fidelity \cite{uhlmann_transition_1985}, which we define according to the convention
\begin{equation}
	F(\rho, \sigma) = \tr\sqrt{\sqrt{\sigma} \rho \sqrt{\sigma}},
\end{equation}
does not satisfy the data processing inequality with the sign given in equation \eqref{eq:DPI}; rather its negative $-F(\rho, \sigma)$ does. However, the proof of theorem \ref{thm:main-theorem} does not actually depend on the sign of the data processing inequality, so the theorem still applies. (The function $f_{\Lambda}$ defined in the preamble to theorem \ref{thm:main-theorem} is at either a minimum or a maximum when the DPI is saturated, depending on the sign of the inequality. In either case, its derivative vanishes and the theorem holds.)

The fidelity is symmetric, so we may compute the derivative with respect to one argument, and then obtain the other by interchanging $\rho $ with $\sigma $. 
We write the fidelity as a function of its first argument as
\begin{equation} \label{eq:fidelity}
	F|_{\rho, \sigma}
		= \left(\tr \circ g_{1/2} \circ \chi_{\sigma^{1/2}}\right)(\rho), 
\end{equation}
where $\chi_{\sigma^{1/2}}$ denotes the conjugation $\chi_{\sigma^{1/2}}(A) = \sigma^{1/2} A \sigma^{1/2}$, and $g_{1/2}$ denotes the square root function $g_{1/2}(A)=A^{1/2}$.
The trace and conjugation maps are linear, so their derivatives can be computed using equation \eqref{eq:linear-derivative}.
The square-root map $g_{1/2}$ is locally analytic, so its derivative can be computed using equation \eqref{eq:locally-analytic-formula}.
Taking the derivative of equation \eqref{eq:fidelity} using the chain rule, simplifying with these formulas, and exploiting the cyclicity of the trace, we obtain the formula
\begin{equation}
	\d_{(1)} F|_{\rho, \sigma}(M)
		= \frac{1}{2} \tr \left( \sigma^{1/2} (\sigma^{1/2} \rho \sigma^{1/2})^{-1/2} \sigma^{1/2} M \right).
\end{equation}
Dualizing according to equation \eqref{eq:dualize} gives the gradient of $F$ as
\begin{equation}
	\Del_{(1)} F|_{\rho, \sigma}(M)
	= \frac{1}{2} \sigma^{1/2} (\sigma^{1/2} \rho \sigma^{1/2})^{-1/2} \sigma^{1/2},
\end{equation}
a result previously shown in \cite{CGW}.
So by theorem \ref{thm:main-theorem}, saturation of the data processing inequality for the fidelity implies the equation
\begin{equation}
	\sigma^{1/2} (\sigma^{1/2} \rho \sigma^{1/2})^{-1/2} \sigma^{1/2}
	= \Lambda^*\left[ \Lambda(\sigma)^{1/2} \left(\Lambda(\sigma)^{1/2} \Lambda(\rho) \Lambda(\sigma)^{1/2}\right)^{-1/2} \Lambda(\sigma)^{1/2}\right].
\end{equation}
The ``second gradient'' version of this equation is the same, but with $\rho $ and $\sigma $ interchanged.
Each equation individually implies DPI saturation, which can be seen by a similar calculation to that given for the relative entropy in the previous example.

\subsubsection{Sandwiched R\'enyi Relative Entropy} \label{subsubsec:alpha-renyis}
The sandwiched R\'{e}nyi relative entropies \cite{wilde_strong_2014, muller-lennert_quantum_2013} are a family of distinguishability measures defined by 
\begin{equation}
	\tilde{D}_\alpha (\rho ,\sigma ) = \frac{1}{\alpha -1}  \log \tr \left[ \left( \sigma ^{\gamma } \rho \sigma ^{\gamma } \right)^\alpha   \right]
\end{equation}
with $\gamma \defi \frac{1-\alpha }{2\alpha }$. They are known to satisfy the data-processing inequality for $\alpha \in [1/2, \infty)$ \cite{muller-lennert_quantum_2013}.
As a function of $\rho$, $\tilde{D}_{\alpha}$ may be decomposed as
\begin{equation}
	\tilde{D}_{\alpha}|_{\rho, \sigma}(\rho)
		= [L_{1/(\alpha-1)} \circ \log \circ \tr \circ g_{\alpha} \circ \chi_{\sigma^\gamma}](\rho),
\end{equation}
where $L_{1/(\alpha-1)}$ is multiplication by $1/(\alpha-1),$ $g_{\alpha}$ is the power function $g_{\alpha}(A) = A^{\alpha},$ and $\chi_{\sigma^{\gamma}}$ is the conjugation function $\chi_{\sigma^{\gamma}}(A) = \sigma^{\gamma} A \sigma^{\gamma}.$
The maps $L_{1/(\alpha-1)}$ and $\log$ are functions of real numbers, so their derivatives are standard.
The trace and conjugation maps are linear, so their derivatives can be computed using equation \eqref{eq:linear-derivative}.
The power map $g_{\alpha}$ is locally analytic, so its derivative can be computed using equation \eqref{eq:locally-analytic-formula}.
Applying the chain rule, simplifying with these formulas, and exploiting the cyclic property of the trace, we obtain the formula
\begin{equation}
	\d_{(1)} \tilde{D}_{\alpha}|_{\rho, \sigma}(M)
		= \frac{\alpha}{\alpha-1} \frac{1}{\tr[(\sigma^{\gamma} \rho \sigma^{\gamma})^{\alpha}]}
			\tr \left[ \sigma^{\gamma} (\sigma^{\gamma} \rho \sigma^{\gamma})^{\alpha - 1} \sigma^{\gamma} M \right].
\end{equation}
From this expression we may read off the gradient as
\begin{align} \label{eq:sandwiched-first-gradient}
	\grad _{(1)} \tilde{D}_\alpha |_{\rho ,\sigma } &= \frac{\alpha }{\alpha -1} \frac{1}{\tr \left[ \left( \sigma ^{\gamma } \rho \sigma ^{\gamma } \right)^\alpha   \right]} \sigma ^{\gamma } \left( \sigma ^{\gamma } \rho  \sigma ^{\gamma } \right)^{\alpha -1} \sigma ^\gamma.
\end{align}
The gradient with respect to the second argument is more complicated; we delay the formula to the following subsection where we present it as a special case of the $\alpha$-$z$ R\'{e}nyi relative entropies.

Applying \cref{thm:main-theorem} to the gradient from equation \eqref{eq:sandwiched-first-gradient} gives
\begin{align}
	\sigma ^{\gamma } \left( \sigma ^{\gamma } \rho  \sigma ^{\gamma } \right)^{\alpha -1} \sigma ^\gamma &=   \Lambda (\sigma ) ^{\gamma } \left( \Lambda (\sigma ) ^{\gamma } \Lambda (\rho )  \Lambda (\sigma ) ^{\gamma } \right)^{\alpha -1} \Lambda (\sigma ) ^\gamma,
	\label{eq:SRRE_condition}
\end{align}
where we have also made use of $\tilde{D}_\alpha (\rho ,\sigma ) = \tilde{D}_\alpha \left(\Lambda (\rho ),\Lambda (\sigma )\right)$ directly to deal with the trace term in the denominator of \eqref{eq:sandwiched-first-gradient}. 
The same equation was originally derived in \cite{leditzky_data_2016}, and recently rederived in \cite{wang_revisiting_2020} using different techniques.

\subsubsection{$\alpha$-$z$ R\'enyi relative entropy} \label{subsubsec:alpha-z}
The $\alpha$-$z$ R\'{e}nyi relative entropies \cite{audenaert_alpha-z-relative_2015} generalize the sandwiched R\'{e}nyi relative entropies. They are defined by the equation 
\begin{equation}
	D_{\alpha ,z} (\rho ,\sigma ) = \frac{1}{\alpha -1}  \log \tr \left[ \left( \sigma ^{\gamma } \rho^{\frac{\alpha }{z}} \sigma ^{\gamma } \right)^z   \right]
\end{equation}
with $\gamma \defi \frac{1-\alpha }{2z}$.
The $\alpha$-$z$ R\'enyi relative entropies satisfy the DPI for the following ranges of parameters \cite{zhang_wigner-yanase-dyson_2020}:
\begin{itemize}
	\item $0<\alpha <1$ and $z\geq \textrm{max}\left\{ \alpha ,1-\alpha  \right\}$;
	\item $1<\alpha \leq 2$ and $\frac{\alpha }{2} \leq z \leq \alpha $; and
	\item $2\leq \alpha  < \infty $ and $\alpha -1 \leq z \leq \alpha $.
\end{itemize}
For $\alpha =z$, this reduces to the sandwiched R\'enyi relative entropy, and the DPI range agrees with the one from section \ref{subsubsec:alpha-renyis}.
The first and second gradients of $D_{\alpha, z}$ can be computed by decomposing $D_{\alpha, z}$ as
\begin{equation} \label{eq:first-alpha-z}
	D_{\alpha, z}|_{\rho, \sigma}(\rho) = [L_{1/(\alpha - 1)} \circ \log \circ \tr \circ g_{z} \circ \chi_{\sigma^{\gamma}} \circ g_{\alpha/z}](\rho)
\end{equation}
and
\begin{equation} \label{eq:second-alpha-z}
	D_{\alpha, z}|_{\rho, \sigma}(\sigma) = [L_{1/(\alpha - 1)} \circ \log \circ \tr \circ g_{z} \circ (g_{\gamma} \cdot (L_{\rho^{\alpha/z}} \circ g_{\gamma}))](\sigma).
\end{equation}
As in the previous subsection, $L$ denotes left multiplication, $\chi$ denotes conjugation, and $g$ denotes the power function; the derivatives of all these functions can be computed using equations \eqref{eq:linear-derivative} (for $L, \chi,$ and $\tr$) and \eqref{eq:locally-analytic-formula} (for $\log$ and $g$).
Applying the chain rule to compute the derivative of \eqref{eq:first-alpha-z}, we obtain the expression
\begin{align}
	\d_{(1)} D_{\alpha, z}|_{\rho, \sigma}(M)
		& = \frac{1}{\alpha - 1} \frac{1}{\tr[(\sigma^{\gamma} \rho^{\alpha/z} \sigma^{\gamma})^{z}]}
			\nonumber \\
		& \quad \tr\left[ \sum_j z \lambda_j^{z-1} \Lambda_j \sigma^{\gamma} \d g_{\alpha/z}|_{\rho}(M) \sigma^{\gamma} \Lambda_j + \sum_{j \neq k} \frac{\lambda_j^{z} - \lambda_k^z}{\lambda_j - \lambda_k} \Lambda_j \d g_{\alpha/z}|_{\rho}(M) \Lambda_k \right],
\end{align}
where $\sum_j \lambda_j \Lambda_j$ is the spectral decomposition of $\sigma^{\gamma} \rho^{\frac{\alpha}{z}} \sigma^{\gamma},$ and we have not yet explicitly written out the final chain rule term $\d g_{\alpha/z}|_{\rho}(M).$ Applying the cyclic property of the trace, we find that the second term vanishes, and the remaining expression simplifies to
\begin{align}
	\d_{(1)} D_{\alpha, z}|_{\rho, \sigma}(M)
	& = \frac{1}{\alpha - 1} \frac{1}{\tr[(\sigma^{\gamma} \rho^{\alpha/z} \sigma^{\gamma})^{z}]}
	\nonumber \\
	& \quad \tr\left[ z \sigma^{\gamma} (\sigma^{\gamma} \rho^{\frac{\alpha}{z}} \sigma^{\gamma})^{z-1} \sigma^{\gamma} \d g_{\alpha/z}|_{\rho}(M) \right].
\end{align}
If we write the spectral decomposition of $\rho$ as $\sum_j p_j \Pi_j,$ then we have
\begin{equation}
	\d g_{\alpha/z}|_{\rho}(M)
		= \sum_j \frac{\alpha}{z} p_j^{\alpha/z - 1} \Pi_j M \Pi_j
			+ \sum_{j \neq k} \frac{p_j^{\alpha/z} - p_k^{\alpha/z}}{p_j - p_k} \Pi_j M \Pi_k.
\end{equation}
Inserting this into the previous equation and exploiting the cyclic property of the trace, we obtain
\begin{align}
	\d_{(1)} D_{\alpha, z}|_{\rho, \sigma}(M)
	& = \frac{z}{\alpha - 1} \frac{1}{\tr[(\sigma^{\gamma} \rho^{\alpha/z} \sigma^{\gamma})^{z}]}
		\tr\left[ \sum_j \frac{\alpha}{z} p_j^{\alpha/z - 1} \Pi_j \sigma^{\gamma} (\sigma^{\gamma} \rho^{\frac{\alpha}{z}} \sigma^{\gamma})^{z-1} \sigma^{\gamma} \Pi_j M \right. \nonumber \\
	& \quad \left.
	+ \sum_{j \neq k} \frac{p_j^{\alpha/z} - p_k^{\alpha/z}}{p_j - p_k} \Pi_k \sigma^{\gamma} (\sigma^{\gamma} \rho^{\frac{\alpha}{z}} \sigma^{\gamma})^{z-1} \sigma^{\gamma} \Pi_j M\right].
\end{align}
A completely analogous calculation for the derivative of \eqref{eq:second-alpha-z} gives
\begin{align}
	\d_{(2)} D_{\alpha ,z} |_{\rho ,\sigma }(M) & = \frac{z}{\alpha-1} \frac{1}{\tr \left[ \left( \sigma ^{\gamma } \rho^{\frac{\alpha }{z}} \sigma ^{\gamma } \right)^z   \right]} \bigg[ \sum_{j}^{} \gamma \mu_j^{\gamma -1} \Phi_j \left\{ \left( \sigma ^\gamma \rho^{\frac{\alpha }{z}} \sigma ^\gamma  \right)^z , \sigma ^{-\gamma } \right\} \Phi_j M \\ 
	&\ + \sum_{j\neq k}^{} \frac{\mu  _j^\gamma - \mu  _k^\gamma  }{\mu  _j - \mu  _k  } \Phi_k \left\{ \left( \sigma ^\gamma \rho^{\frac{\alpha }{z}} \sigma ^\gamma  \right)^z , \sigma ^{-\gamma } \right\} \Phi_j M \bigg],
\end{align}
where $\sum_j \mu_j \Phi_j$ is the spectral decomposition of $\sigma$ and $\{A, B\} = A B + B A$ is the anticommutator.
From these two expressions, we may immediately read off the gradients as
\begin{align} \label{eq:alpha-z-stuff-1}
	\grad _{(1)} D_{\alpha ,z} |_{\rho ,\sigma } &=  \frac{z}{\alpha -1} \frac{1}{\tr \left[ \left( \sigma ^{\gamma } \rho^{\frac{\alpha }{z}} \sigma ^{\gamma } \right)^z   \right]} \bigg[ \sum_{j}^{}\frac{\alpha }{z} \lambda _j^{\frac{\alpha -z }{z}} \Pi_j \sigma ^{\gamma } \left( \sigma ^{\gamma } \rho^{\frac{\alpha }{z}}  \sigma ^{\gamma } \right)^{z -1} \sigma ^\gamma \Pi_j \\  &\ +  \sum_{j \neq k}^{}\frac{\lambda  _j^{\frac{\alpha }{z}} - \lambda _k^{\frac{\alpha }{z}}  }{\lambda _j - \lambda _k } \Pi_k \sigma ^{\gamma } \left( \sigma ^{\gamma } \rho^{\frac{\alpha }{z}}  \sigma ^{\gamma } \right)^{z -1} \sigma ^\gamma \Pi_j \bigg], \\ \label{eq:alpha-z-stuff-2}
	\grad _{(2)} D_{\alpha ,z} |_{\rho ,\sigma } &= \frac{z}{\alpha-1} \frac{1}{\tr \left[ \left( \sigma ^{\gamma } \rho^{\frac{\alpha }{z}} \sigma ^{\gamma } \right)^z   \right]} \bigg[ \sum_{j}^{} \gamma \mu_j^{\gamma -1} \Phi_j \left\{ \left( \sigma ^\gamma \rho^{\frac{\alpha }{z}} \sigma ^\gamma  \right)^z , \sigma ^{-\gamma } \right\} \Phi_j \\ 
	&\ + \sum_{j\neq k}^{} \frac{\mu  _j^\gamma - \mu  _k^\gamma  }{\mu  _j - \mu  _k  } \Phi_k \left\{ \left( \sigma ^\gamma \rho^{\frac{\alpha }{z}} \sigma ^\gamma  \right)^z , \sigma ^{-\gamma } \right\} \Phi_j \bigg].
\end{align}
Substituting these equations into equations \eqref{eq:first-vanishing-gradient} and \eqref{eq:second-vanishing-gradient} from \cref{thm:main-theorem} yields an operator equality; admittedly, this equality is rather complicated in the general case $\alpha \neq z.$ 
However, for $\alpha=z$, it is straightforward to check that the first-gradient condition coincides with equation \eqref{eq:SRRE_condition}.

Both operator equations are sufficient to imply DPI saturation, which follows from the remark after theorem \ref{thm:main-theorem}, because the $\alpha $-$z$ entropy varies under scalar multiplication as
\begin{align}
	D_{\alpha ,z} (k \rho , k' \sigma ) = D_{\alpha ,z} (\rho ,\sigma ) + \frac{\alpha}{\alpha-1} \log k - \log k'.
\end{align}
This is clearly an invertible function of $D_{\alpha ,z}(\rho ,\sigma )$.

This result seems to be distinct from known operator equalities from two previous works.
The first of these equations was derived in \cite{chehade_saturating_2020}, and is given by
\begin{align}
	f(\rho ,\sigma ) &=  \Lambda ^{\dagger}(f(\Lambda (\rho ),\Lambda (\sigma ))), \label{eq:f} \\
	\text{with } f(\rho ,\sigma ) &=  \sigma ^{\frac{1-z}{2z}} \left( \sigma ^{\frac{1-\alpha }{2z}} \rho^{\frac{\alpha }{z}}  \sigma ^{\frac{1-\alpha }{2z}}  \right)^{z-1} \sigma ^{\frac{1-z}{2z}}.
\end{align}
This result is necessary and sufficient for DPI saturation, meaning it must be equivalent to our equations; however, this equivalence is not obvious by inspection.
Secondly, \cite{zhang_equality_2020} shows a similar result where instead $f$ is defined to be
\begin{align}
	f(\rho , \sigma ) &=  \sigma ^{ \frac{1-\alpha }{2z} } \left(\sigma ^{ \frac{1-\alpha }{2z} } \rho ^{\frac{\alpha }{z}} \sigma ^{ \frac{1-\alpha }{2z} }  \right)^{\alpha -1} \sigma ^{ \frac{1-\alpha }{2z} } .
\end{align}

\subsubsection{Quantum $f$-divergences}
For a real-valued differentiable function $f$ on $(0,\infty )$, the quantum $f$-divergence is defined by \cite{petz_quasi-entropies_1985}
\begin{align}
	D_f (\rho  \| \sigma ) \defi \tr\left( \sqrt{\sigma } f(L_\rho  R_\sigma ^{-1}) (\sqrt{\sigma }) \right),
\end{align}
where the superoperator $L_\rho $ acts on matrices via left multiplication by $\rho $ and similarly $R_\sigma $ acts via right multiplication by $\sigma $.
This recovers the relative entropy for $f(x) = x \log x$, and the Petz-R\'enyi relative entropy for $f(x) = x^\alpha $.
From \cite{hiai_quantum_2011} (theorem 5.1), it is known that DPI is satisfied for any operator convex function $f$ on $[0,\infty )$.

With spectral decompositions $\rho =\sum_{j}^{} p_j \Pi_j$ and $\sigma  =\sum_{k}^{} \mu_k \Phi_k$, and assuming both matrices are invertible, we have
\begin{align}
	D_f(\rho \|\sigma ) &= \sum_{j,k}^{} \mu_k f(p_j/\mu_k) \tr(\Pi_j \Phi_k),
\end{align}
which we can also write asymetrically as
\begin{equation}
	D_{f,\sigma}(\rho) = \sum_{k} \mu_k \tr\left( f(\rho / \mu_k) \Phi_k \right).
\end{equation}
Since the dependence of this function on $f(\rho / \mu_k)$ is linear, its derivative with respect to $\rho$ is obtained by replacing $f(\rho / \mu_k)$ with that function's derivative with respect to $\rho$. Assuming that $f$ is locally analytic at $\rho / \mu_k$, we obtain
\begin{align}
	\d D_{f,\sigma}|_{\rho}(M)
		= \sum_{k} & \tr\left( \sum_{j} f'(p_j / \mu_k) \Pi_j \Phi_k \Pi_j M 
				\right. \\
				& \quad \left. + \sum_{j \neq j'} \mu_k \frac{f(p_j / \mu_k) - f(p_{j'}/\mu_k)}{p_j - p_j'} \Pi_{j'} \Phi_k \Pi_j M \right).
\end{align}
From this we can read off the gradient of $D_f$ with respect to the first argument as
\begin{align}
	\grad _{(1)} D_f(\rho  \|\sigma )
		= \sum_{k}& \left( \sum_{j} f'(p_j / \mu_k) \Pi_j \Phi_k \Pi_j 
		\right. \\
		& \quad \left. + \sum_{j \neq j'} \mu_k \frac{f(p_j / \mu_k) - f(p_{j'}/\mu_k)}{p_j - p_j'} \Pi_{j'} \Phi_k \Pi_j \right),
\end{align}
giving us a new operator equation again according to \cref{thm:main-theorem}. If we set $f(x) = x \log x$, then it is straightforward to check that this expression is equivalent to equation \eqref{eq:rel-ent-grad} for the gradient of relative entropy. The gradient with respect to the second argument is more complicated so we omit the expression, but it can be derived using the same technique.

In addition to the examples given here, similar expressions can be derived for the maximal $f$-divergences \cite{matsumoto_new_2018}, the Jencova-Ruskai function \cite{jencova_unified_2010}, and any other smooth distinguishability measures.

\section{Saturation of DPI for positive semidefinite matrices} \label{sec:boundary}

In this section, we generalize the results of section \ref{sec:main} to distinguishability measures $\mathcal{B}$ that are defined not just on the space of positive matrices, but on the space of positive semidefinite matrices. The most general result is less aesthetically simple than the one given in theorem \ref{thm:main-theorem}, but is still broadly applicable. The basic idea is that while equation \eqref{eq:first-vanishing-gradient} need not hold as a general operator equation when one of the matrices $\rho$ or $\sigma$ has a vanishing eigenvalue, it does hold in the orthogonal complement of the space of operators that act on the zero-eigenspace of $\rho$.

In the first subsection we introduce the geometry of the space of positive semidefinite operators $\psd(\mathcal{H}).$ In the second, we prove a generalization of theorem \ref{thm:main-theorem} that holds for matrices with vanishing eigenvalues. In the third, we work out an explicit example of this theorem: an operator equation implied by DPI saturation for the relative entropy between two density matrices when one has a nontrivial zero-eigenspace. We then compare our equation to a similar equation from \cite{hiai_quantum_2011}.

\subsection{Geometry of the positive semidefinite space}

As discussed in section \ref{sec:matrix-manifolds}, the Hermitian operators on an $n$-dimensional Hilbert space $\mathcal{H}$ form an $n^2$-dimensional real manifold $\herm(\mathcal{H}).$ The set of positive operators, $\pos(\mathcal{H})$, is an $n^2$-dimensional real submanifold. Since $\pos(\mathcal{H})$ consists of all Hermitian operators whose eigenvalues are strictly positive, its topological closure consists of all Hermitian operators whose eigenvalues are nonnegative -- we denote this space of positive semidefinite operators by $\psd(\mathcal{H}).$ The \emph{boundary} $\del \psd(\mathcal{H})$ consists of all operators that are in $\psd(\mathcal{H})$ but not in $\pos(\mathcal{H})$; i.e., those with at least one vanishing eigenvalue.

For an operator $\rho$ in $\psd(\mathcal{H})$, on the boundary or otherwise, the tangent space $T_{\rho} \psd(\mathcal{H})$ consists of the Hermitian operators that are orthogonal to all operators acting on the zero-eigenspace of $\rho $, with respect to the Hilbert-Schmidt inner product.
In other words, if we think of $\herm(\mathcal{H})$ as a Hilbert space with respect to the Hilbert-Schmidt inner product, and denote the zero-eigenspace of $\rho$ by $E_0(\rho)$, then $T_{\rho} \psd(\mathcal{H})$ is the orthogonal complement of $\herm(E_0(\rho))$ within $\herm(\mathcal{H})$.
We provide a simple proof of this fact in \cref{app:psd}.
If $k$ is the dimension of $E_0(\rho)$, then $T_{\rho} \psd(\mathcal{H})$ has real dimension $n^2-k^2$.

\subsection{Data processing equality for positive semidefinite matrices}

Let $\mathcal{B}$ be a distinguishability measure such that at least one argument can be taken to be positive \emph{semi}definite, i.e.,
\begin{equation}
	\mathcal{B} : \psd(\mathcal{H}) \times \pos(\mathcal{H}) \rightarrow \reals.
\end{equation}
If we assume that $\mathcal{B}$ is continuous in the limit as its first argument is taken to the boundary $\del \psd(\mathcal{H}),$ then the data processing inequality holds in that limit by continuity as well.

Now, let $\rho$ and $\sigma$ be density matrices, and $\Lambda$ a linear map of Hermitian operators, such that the data processing inequality
\begin{equation}
	\mathcal{B}(\rho, \sigma) \geq \mathcal{B}(\Lambda(\rho), \Lambda(\sigma))
\end{equation}
holds in a neighborhood of $\rho$ in $\psd(\mathcal{H}).$ (Cf. equation \eqref{eq:DPI-neighborhood}.) Then, by the same arguments given in section \ref{sec:main}, the function
\begin{equation} \label{eq:general-f}
	f_{\sigma, \Lambda}(\rho) = \mathcal{B}(\rho, \sigma) - \mathcal{B}(\Lambda(\rho), \Lambda(\sigma))
\end{equation}
must be locally minimal \emph{in any direction tangent to $\psd(\mathcal{H}).$} It need not be locally minimal in directions non-tangent to $\psd(\mathcal{H})$, since the DPI need not be satisfied outside of $\psd(\mathcal{H})$ and $f_{\sigma, \Lambda}$ may go negative. So the equation $\d f_{\sigma, \Lambda} |_{\rho} = 0$, which was essential to our analysis in section \ref{sec:main}, need not be satisfied --- or even defined --- for all vectors tangent to $\rho.$

What we have instead is the following condition: for every $M \in T_{\rho}\psd(\mathcal{H})$, the derivative
\begin{equation}
	\d f_{\sigma, \Lambda}|_{\rho}(M) = \lim_{\epsilon \rightarrow 0} \frac{f_{\sigma, \Lambda}(\rho + \epsilon M) - f_{\sigma, \Lambda}(\rho)}{\epsilon}
\end{equation}
exists and vanishes.
In terms of the explicit form of $f_{\sigma, \Lambda}$ written in equation \eqref{eq:general-f}, this condition can be written using the chain rule as
\begin{equation} \label{eq:derivative-equality-psd-prelim}
	\d \mathcal{B}_{\sigma}|_{\rho}(M) = \d \mathcal{B}_{\Lambda(\sigma)}|_{\Lambda(\rho)}(\Lambda(M)),
\end{equation}
where we have introduced the notation $\mathcal{B}_{\sigma}(\bullet) \equiv \mathcal{B}(\bullet, \sigma).$
There is one potential issue with this expression, which is that we have not yet determined whether $\d \mathcal{B}_{\sigma}|_{\rho}$ being defined on $M$ implies that $\d \mathcal{B}_{\Lambda(\sigma)}|_{\Lambda(\rho)}$ is defined on $\Lambda(M)$.
However, this follows from a fairly straightforward observation --- $M$ is a tangent vector at $\rho$ by definition when $\rho + \epsilon M$ is positive semidefinite for sufficiently small $\epsilon$; because $\Lambda$ is linear and sends positive semidefinite operators to positive semidefinite operators, this implies that $\Lambda(\rho) + \epsilon \Lambda(M) = \Lambda(\rho + \epsilon M)$ is positive semidefinite for sufficiently small $\epsilon,$ so $M$ being in $T_{\rho}\psd(\mathcal{H})$ implies that $\Lambda(M)$ is in $T_{\Lambda(\rho)}\psd(\mathcal{H}).$

Equation \eqref{eq:derivative-equality-psd-prelim} is an equality of linear functionals of $T_{\rho} \psd(\mathcal{H}).$
This is already a useful equation in practice, since it gives $\dim(T_{\rho}\psd(\mathcal{H})) = n^2 - \rank(E_0(\rho))^2$ scalar equations constraining $\rho$ and $\sigma$ from the single scalar equation $f_{\sigma, \Lambda}(\rho) = 0.$
However, for aesthetic purposes, it is nice to be able to write equation \eqref{eq:derivative-equality-psd-prelim} not as an equality of functionals, but as an equality of operators.
To do this, we first make the equality of linear functionals on $T_{\rho} \psd(\mathcal{H})$ onto an equality of linear functionals on all of $\herm(\mathcal{H})$ by precomposing with the projection onto $T_{\rho} \psd(\mathcal{H}).$
A convenient way to write this expression is to note that the zeroth power $\rho^{0}$ is the orthogonal projection onto the nonzero eigenspaces of $\rho$, and $(1 - \rho^{0})$ the orthogonal projection onto its zero eigenspace.
The space $T_{\rho} \psd(\mathcal{H})$ can then be written as
\begin{equation}
	T_{\rho} \psd(\mathcal{H}) = \{ M \in \herm(\mathcal{H}) | (1 - \rho^0) M (1 - \rho^0) = 0 \},
\end{equation}
and the projection of an arbitrary $M \in \herm(\mathcal{H})$ onto $T_{\rho} \psd(\mathcal{H})$ is $M - (1 - \rho^0) M (1 - \rho^0).$
Equation \eqref{eq:derivative-equality-psd-prelim} is then equivalent to the statement that for arbitrary $M \in \herm(\mathcal{H}),$ we have
\begin{equation} \label{eq:extended-functional}
	\d \mathcal{B}_{\sigma}|_{\rho}(M - (1 - \rho^{0}) M (1 - \rho^{0})) = \d \mathcal{B}_{\Lambda(\sigma)}|_{\Lambda(\rho)}(\Lambda(M - (1 - \rho^{0})M(1 - \rho^{0}))).
\end{equation}
Since $G_{\rho, \sigma}(M) = \d \mathcal{B}_{\sigma}|_{\rho}(M - (1 - \rho^0) M (1 - \rho^0))$ is a linear functional on $\herm(\mathcal{H}),$ it can be dualized using the Hilbert-Schmidt inner product to construct a unique operator $\Del \mathcal{B}_{\sigma}|_{\rho}$ satisfying
\begin{equation} \label{eq:zero-gradients}
	\tr(\Del \mathcal{B}_{\sigma}|_{\rho} M) = G_{\rho, \sigma}(M).
\end{equation}

Let us now momentarily denote by $Q$ the expression $M - (1 - \rho^0) M (1 - \rho^0).$
Since we have
\begin{equation}
	(1 - \Lambda(\rho)^0) \Lambda(Q) (1 - \Lambda(\rho)^0)= 0,
\end{equation}
which is just the statement we made before that  $\Lambda(Q)$ is tangent to $\Lambda(\rho)$ because $Q$ is tangent to $\rho,$ we may write
\begin{equation}
	\d \mathcal{B}_{\Lambda(\sigma)}|_{\Lambda(\rho)}(\Lambda(Q))
		= \d \mathcal{B}_{\Lambda(\sigma)}|_{\Lambda(\rho)}(\Lambda(Q) - (1 - \Lambda(\rho)^0) \Lambda(Q) (1 - \Lambda(\rho)^0)) = G_{\Lambda(\rho), \Lambda(\sigma)}(\Lambda(Q)),
\end{equation}
and hence as
\begin{equation}
	\d \mathcal{B}_{\Lambda(\sigma)}|_{\Lambda(\rho)}(\Lambda(Q))
	= \tr( \Del \mathcal{B}_{\Lambda(\sigma)}|_{\Lambda(\rho)} \Lambda(Q)).
\end{equation}
By taking the dual of the map $\Lambda$ appearing on the right-hand side of this expression, re-inserting $Q = M - (1 - \rho^0) M (1 - \rho^0),$ and exploiting the cyclic property of the trace, we obtain
\begin{equation}
	\d \mathcal{B}_{\Lambda(\sigma)}|_{\Lambda(\rho)}(\Lambda(Q))
	= \tr( \Lambda^*(\Del \mathcal{B}_{\Lambda(\sigma)}|_{\Lambda(\rho)}) M)
		- \tr( (1 - \rho^0) \Lambda^*(\Del \mathcal{B}_{\Lambda(\sigma)}|_{\Lambda(\rho)}) (1 - \rho^0) M).
\end{equation}
Plugging this back into equation \eqref{eq:extended-functional} gives us
\begin{equation}
	\tr(\Del \mathcal{B}_{\sigma}|_{\rho} M) = \tr( \Lambda^*(\Del \mathcal{B}_{\Lambda(\sigma)}|_{\Lambda(\rho)}) M)
- \tr( (1 - \rho^0) \Lambda^*(\Del \mathcal{B}_{\Lambda(\sigma)}|_{\Lambda(\rho)}) (1 - \rho^0) M).
\end{equation}
And finally, since this equation must hold for arbitrary $M \in \herm(\mathcal{H}),$ it implies the operator equation
\begin{equation} \label{eq:zero-gradient-final-equality}
	\Del \mathcal{B}_{\sigma}|_{\rho}
		 = \Lambda^*(\Del \mathcal{B}_{\Lambda(\sigma)}|_{\Lambda(\rho)})
	- (1 - \rho^0) \Lambda^*(\Del \mathcal{B}_{\Lambda(\sigma)}|_{\Lambda(\rho)}) (1 - \rho^0).
\end{equation}
In words, what this equation tells us is the following: the gradient of $\mathcal{B}_{\sigma}$ at $\rho,$ defined as an element of $\herm(\mathcal{H})$ by extending $\d \mathcal{B}_{\sigma}$ orthogonally to $T_{\rho} \psd(\mathcal{H})$ using the zero map, equals the $\Lambda^*$-image of the gradient of $\mathcal{B}_{\Lambda(\sigma)}$ at $\Lambda(\rho),$ \emph{outside} of the subspace whose orthogonal projector is $(1 - \rho^0).$

For $\rho \in \pos(\mathcal{H})$, this gives \cref{thm:main-theorem} as a special case, since we have $T_{\rho }\psd(\mathcal{H}) \cong \herm(\mathcal{H})$ and $\rho^0 = 1.$

\subsection{Example: relative entropy}

Let $\mathcal{H}$ be an $n$-dimensional Hilbert space, and let $\rho$ be a density matrix on $\mathcal{H}$ with $k>0$ vanishing eigenvalues.
Let $\sigma$ be an arbitrary positive operator on $\mathcal{H}.$
If $\rho$ were positive, then we could apply the result of section \ref{subsubsec:relative-entropy} to show that the equation $D(\rho||\sigma) = D(\Lambda(\rho)||\Lambda(\sigma))$ implies
\begin{equation} \label{eq:rel-ent-boundary-matrix}
	\log(\rho) - \log(\sigma) = \Lambda^*[\log(\Lambda(\rho)) - \log(\Lambda(\sigma))].
\end{equation}

Since $\rho$ is only positive \emph{semi}definite, we must apply the results of the previous subsection.
The first step is to define a map $\log^{\times}$ on positive semidefinite operators that acts as $\log$ on the nonzero eigenvalues, but keeps the zero eigenvalues zero.
The relative entropy is then defined as
\begin{equation}
	D(\rho || \sigma) = \tr(\rho \log^{\times}(\rho) - \rho \log(\sigma)).
\end{equation}
This extension of the relative entropy from $\pos(\mathcal{H}) \times \pos(\mathcal{H})$ to $\psd(\mathcal{H}) \times \pos(\mathcal{H})$ still satisfies the data processing inequality.
Its $\rho$-derivative acting on operators $M$ in $T_{\rho}\psd(\mathcal{H})$ can be computed straightforwardly using the techniques of section \ref{sec:matrix-manifolds}, and is given by
\begin{equation}
	\d D_{\sigma}|_{\rho}(M) = \tr(\log^{\times}(\rho) M - \log(\sigma) M + \rho^0 M).
\end{equation}
We extend this to act on arbitrary Hermitian operators $M$ by precomposing with the projection onto $T_{\rho}\psd(\mathcal{H}),$ giving
\begin{equation}
	G_{\rho, \sigma}(M) = \tr(\log^{\times}(\rho) M - \log(\sigma) M + (1 - \rho^0) \log(\sigma) (1 - \rho^0) M + \rho^0 M),
\end{equation}
where we have used the identity $\log^{\times}(\rho) (1 - \rho^0) = \rho^0 (1 - \rho^0) = 0.$
It is clear from this expression and equation \eqref{eq:zero-gradients} that the gradient of $D_{\sigma}$ at $\rho$ is given by
\begin{equation}
	\Del D_{\sigma}|_{\rho}
		= \log^{\times}(\rho) - \log(\sigma) + (1 - \rho^0) \log(\sigma) (1 - \rho^0) + \rho^0.
\end{equation}
The operator equation \eqref{eq:zero-gradient-final-equality} implied by DPI saturation, for this particular case, can then be written as
\begin{align} \label{eq:zeros-log-prelim}
	& \log^{\times}(\rho) - \log(\sigma)_{E_0(\rho)^{\perp}} + \rho^0 \nonumber \\
	& \quad = \Lambda^*(\log^{\times}(\Lambda(\rho)) - \log(\Lambda(\sigma))_{E_0(\Lambda(\rho))^{\perp}} + \Lambda(\rho)^0)_{E_0(\rho)^{\perp}},
\end{align}
where we have introduced the notation
\begin{equation}
	A_{E_0(\rho)^{\perp}}
		= A - (1 - \rho^0) A (1 - \rho^0).
\end{equation}

This expression can be simplified by using lemma 3.2(ii) of \cite{hiai_quantum_2011}, which implies that since $\Lambda$ is a CPTP map it satisfies
\begin{equation}
	\rho^{0} \Lambda^*(\Lambda(\rho)^0) = \Lambda^*(\Lambda(\rho)^0) \rho^{0} = \rho^0.
\end{equation}
This gives us
\begin{equation}
	\Lambda^*(\Lambda(\rho)^0)_{E_0(\rho)^{\perp}}
		= \Lambda^*(\Lambda(\rho)^0) - (1 - \rho^0) \Lambda^*(\Lambda(\rho)^0) (1 - \rho^0)
		= \rho^0,
\end{equation}
which simplifies equation \eqref{eq:zeros-log-prelim} to
\begin{equation} \label{eq:zeros-log}
	\log^{\times}(\rho) - \log(\sigma)_{E_0(\rho)^{\perp}}
		= \Lambda^*(\log^{\times}(\Lambda(\rho)) - \log(\Lambda(\sigma))_{E_0(\Lambda(\rho))^{\perp}})_{E_0(\rho)^{\perp}}.
\end{equation}

Note that equation \eqref{eq:zeros-log} is sufficient to imply DPI saturation for $\rho$ and $\sigma$ even though both sides of the equation vanish on the subspace of operators acting on the zero eigenspace of $\rho$. This follows from the general identity
\begin{equation}
	\tr(\rho A_{E_0(\rho)^{\perp}}) = \tr(\rho A),
\end{equation}
which in turn follows from $\rho (1 - \rho^0) = 0.$ If we multiply equation \eqref{eq:zeros-log} on the left by $\rho$ and take the trace, then we obtain
\begin{equation}
	\tr(\rho \log^{\times}(\rho) - \rho \log(\sigma))
	= \tr(\Lambda(\rho) \log^{\times}(\Lambda(\rho)) - \Lambda(\rho) \log(\Lambda(\sigma))).
\end{equation}

As a final comment, we note that our equation \eqref{eq:zeros-log} is very similar to theorem 5.1(ix) from \cite{hiai_quantum_2011}, which states that saturation of the DPI for relative entropy implies
\begin{equation} \label{eq:hiai-equation}
	\log^{\times} \rho - \log(\sigma) \rho^0 = \Lambda^*(\log^{\times} \Lambda(\rho) - \log(\Lambda(\sigma)) \Lambda(\rho)^0).
\end{equation}
(To obtain this equation from theorem 5.1(ix) of \cite{hiai_quantum_2011}, we have made the notational substitutions $A \mapsto \rho$, $B \mapsto \sigma,$ $\Phi \mapsto \Lambda,$ $\log^{*} \mapsto \log^{\times},$ and used the assumption that $\sigma$ is strictly positive.)
Equation \eqref{eq:hiai-equation} is essentially an asymmetric version of equation \eqref{eq:zeros-log}.
If we multiply \eqref{eq:zeros-log} on the right by $\rho^{0}$, we obtain
\begin{equation} \label{eq:asymmetric-zeros-log}
	\log^{\times}(\rho) - \log(\sigma) \rho^0
	= \Lambda^*(\log^{\times}(\Lambda(\rho)) - \log(\Lambda(\sigma))_{E_0(\Lambda(\rho))^{\perp}}) \rho^0.
\end{equation}
Consistency of equations \eqref{eq:hiai-equation} and \eqref{eq:asymmetric-zeros-log} requires the identity
\begin{equation}
	\Lambda^*(\log^{\times}(\Lambda(\rho)) - \log(\Lambda(\sigma))_{E_0(\Lambda(\rho))^{\perp}}) \rho^0
		= \Lambda^*(\log^{\times}(\Lambda(\rho)) - \log(\Lambda(\sigma)) \Lambda(\rho)^0).
\end{equation}
Curiously, this identity is quite nontrivial to show; for example, the first terms on each side of the equation,
\begin{equation}
	\Lambda^*(\log^{\times}(\Lambda(\rho))) \rho^0 
\end{equation}
and
\begin{equation}
	\Lambda^*(\log^{\times}(\Lambda(\rho))),
\end{equation}
are not equal for general $\rho$ and $\Lambda,$ even though they have no dependence on $\sigma.$
We have not been able to find a direct way to derive equation \eqref{eq:hiai-equation} from equation \eqref{eq:zeros-log}, without using \eqref{eq:zeros-log} to show saturation of the DPI for relative entropy and then using the functional analytic techniques of \cite{hiai_quantum_2011} to derive equation \eqref{eq:hiai-equation}.

\section{Discussion} \label{sec:discussion}

We now comment on some potential directions for future work.

\subsection{Approximate DPI saturation}

Our method gives, for any distinguishability measure $\mathcal{B}$ and any quantum channel $\Lambda$, an operator equation satisfied by $\rho$ and $\sigma$ whenever the data processing inequality is \emph{exactly} saturated. As a reminder to the reader, the essential feature of this argument is that when the DPI is saturated, the function
\begin{equation}
	f_{\Lambda}(\rho, \sigma) = \mathcal{B}(\rho, \sigma) - \mathcal{B}(\Lambda(\rho),\Lambda(\sigma))
\end{equation}
is at its minimum value, and so its gradient with respect to either argument must vanish.

When the DPI is \emph{approximately} saturated, $f_{\Lambda}$ is \emph{close} to its minimum value. There has been a series of breakthroughs in recent years, beginning with \cite{fawzi2015quantum} and culminating in \cite{junge2018universal} (see also \cite{berta2015renyi, zhang2016strengthened}), which showed that smallness of $f_{\Lambda}$ for certain distinguishability measures implies the existence of an approximate recovery channel whose fidelity is controlled by the size of $f_{\Lambda}$. Our formalism may offer an alternate route toward results of this type, as certain distinguishability measures may have the property that $f_{\Lambda}$ being close to its minimum value implies that its gradient is small in an appropriate sense. If this is the case, then approximate DPI saturation still implies an operator equation (smallness of the gradient) that may have applications to approximate quantum error correction.

\textbf{Note added:} In a followup paper \cite{cree2021approximate}, we have undertaken this analysis for the second sandwiched relative R\'{e}nyi entropy, and shown that the geometric formalism developed in the present paper can be used to derive new bounds on the quality of approximate Petz recovery.

\subsection{Connection to Petz recovery}

Saturation of the data processing inequality is famously linked to the existence recovery channels. In \cite{petz1}, Petz showed that saturation of the data processing inequality for the relative entropy for density matrices $\rho$ and $\sigma$ and a quantum channel $\Lambda$ implies the existence of a $\rho$-independent recovery channel $\mathcal{R}_{\sigma}$ satisfying
\begin{align}
	[\mathcal{R}_{\sigma} \circ \Lambda](\rho)
		& = \rho, \\
	[\mathcal{R}_{\sigma} \circ \Lambda](\sigma)
		& = \sigma.
\end{align}
Petz's original proof technique relied heavily on functional analysis. Since in this paper we have used geometric arguments to circumvent functional analytic proofs of certain consequences of DPI saturation, we wonder whether it might be possible to use similar ideas to provide an alternate proof of Petz's theorem without using any functional analysis.

For the moment, we have no concrete suggestions for how this might be done. However, we recall from section \ref{subsubsec:alpha-renyis} that for the sandwiched R\'{e}nyi relative entropies, the operator equation implied by DPI saturation is
\begin{equation}
	\sigma^{\gamma } \left( \sigma^{\gamma} \rho  \sigma^{\gamma } \right)^{\alpha-1} \sigma^\gamma
		= \Lambda^*[\Lambda(\sigma)^{\gamma} \left( \Lambda(\sigma)^{\gamma} \Lambda(\rho)  \Lambda(\sigma)^{\gamma } \right)^{\alpha -1} \Lambda(\sigma)^\gamma].
\end{equation}
For $\alpha = 2$ (i.e., $\gamma = -1/4$), this simplifies to
\begin{equation}
	\sigma^{-1/2} \rho \sigma^{-1/2}
		= \Lambda^*\left[\Lambda(\sigma)^{-1/2} \Lambda(\rho) \Lambda(\sigma)^{-1/2} \right].
\end{equation}
This is exactly the condition Petz originally produced in \cite{petz1}!\footnote{This observation was made previously in \cite{leditzky_data_2016}. We include it for the purpose of discussion, rather than claiming it as an original result.} It tells us that the Petz map
\begin{equation}
	\mathcal{R}_{\sigma}(\bullet)
		= \sigma^{1/2} \Lambda^*\left[\Lambda(\sigma)^{-1/2} \Lambda(\bullet) \Lambda(\sigma)^{-1/2} \right] \sigma^{1/2}
\end{equation}
perfectly recovers $\rho$.

We find it suggestive that the vanishing gradient equation for the sandwiched $2$-R\'{e}nyi relative entropy is exactly equivalent to Petz recovery. While this is enough to prove that DPI saturation for the sandwiched $2$-R\'{e}nyi relative entropy implies recoverability, it is not enough to prove that DPI saturation for the relative entropy implies recoverability. For that, it seems you would still need some functional analytic input telling you that DPI saturation for relative entropy is equivalent to DPI saturation for the sandwiched $2$-R\'{e}nyi relative entropy. However, we remain optimistic that geometric techniques can provide an alternate path to Petz recovery.

Finally, we comment that in \cite{gao_recoverability_2020}, it was shown that approximate DPI saturation implies approximate recoverability for a family of distinguishability measures that includes the sandwiched R\'enyi relative entropies for $\alpha \in [1/2, 1) \cup (1, \infty).$ If it is possible to use our methods to derive operator equations implied by approximate DPI saturation, as mentioned in the previous subsection, it is possible that these techniques could give a new perspective on approximate Petz recovery as well.

\subsection{Nice equations for $\alpha$-$z$ R\'{e}nyi relative entropies}

We find it unsatisfying that while our method produces the exact same equations derived in \cite{leditzky_data_2016} for the sandwiched R\'{e}nyi relative entropies, it does not manifestly produce the same equations for the $\alpha$-$z$ R\'{e}nyi relative entropies that were derived in \cite{chehade_saturating_2020, zhang_equality_2020}. In fact, the equations produced by our method, \eqref{eq:alpha-z-stuff-1} and \eqref{eq:alpha-z-stuff-2}, are significantly more complicated than the ones derived in \cite{chehade_saturating_2020, zhang_equality_2020}. 
However, as mentioned in section \ref{subsubsec:alpha-z}, our equation must be equivalent at least to the equation from \cite{chehade_saturating_2020}, since both are equivalent to DPI saturation.
We remain hopeful that some extra geometric input may make it possible to simplify our equations \eqref{eq:alpha-z-stuff-1} and \eqref{eq:alpha-z-stuff-2} and to understand how it is related to the equations from \cite{chehade_saturating_2020, zhang_equality_2020}.

\ack
We thank Patrick Hayden and Mark Wilde for insightful conversations.
The authors are supported by the Simons Foundation It from Qubit collaboration, by AFOSR (FA9550-16-1-0082), and by DOE Award No.\ DE-SC0019380.

\appendix
\setcounter{section}{0}

\section{Derivatives of locally analytic functions} \label{app:locally-analytic}

As in section \ref{sec:matrix-manifolds}, we will call $f : \herm(\mathcal{H}) \rightarrow \herm(\mathcal{H})$ \emph{locally analytic} at $A \in \herm(\mathcal{H})$ if, in a neighborhood of $A$, $f$ can be written as a Taylor series centered at a multiple of the identity:
\begin{equation}
	f(A) = \sum_{m=0}^{\infty} c_m (A - \alpha I)^{m}.
\end{equation}

To linear order in $\epsilon$, we have
\begin{equation}
	f(A+\epsilon M) = f(A) + \epsilon \sum_{m=0}^{\infty} c_m \sum_{n=0}^{m-1} (A - \alpha I)^{n} M (A - \alpha I)^{m-1-n}.
\end{equation}
Let $\{\lambda_{j}\}$ be the eigenvalues of $A$ and $\{\Pi_{j}\}$ the orthogonal projectors onto its eigenspaces.
Substituting in the spectral decomposition
\begin{equation}
	A - \alpha I
		= \sum_{j} (\lambda_{j} - \alpha) \Pi_{j},
\end{equation}
we find
\begin{equation} \label{eq:locally-analytic-penultimate}
	\d f|_{A}(M)
		= \lim_{\epsilon \rightarrow 0} \frac{f(A + \epsilon M) - f(A)}{\epsilon}
		= \sum_{j, k} \Pi_{j} M \Pi_{k}  \left[\sum_{m=0}^{\infty} c_m \sum_{n=0}^{m-1} (\lambda_j - \alpha)^{n} (\lambda_{k} - \alpha)^{m-1-n}\right].
\end{equation}
For $j \neq k$, we have $\lambda_{j} \neq \lambda_{k}$, and in this case the sum over $n$ in \eqref{eq:locally-analytic-penultimate} simplifies to\footnote{This is just a geometric series in $(\lambda_j - \alpha) / (\lambda_k - \alpha).$}
\begin{equation}
	\sum_{n=0}^{m-1} (\lambda_j - \alpha)^{n} (\lambda_{k} - \alpha)^{m-1-n}
		= \frac{(\lambda_j - \alpha)^m - (\lambda_k - \alpha)^m}{\lambda_j - \lambda_k}.
\end{equation}
Splitting the $j, k$ sum in \eqref{eq:locally-analytic-penultimate} into terms with $j=k$ and terms with $j \neq k$, we have
\begin{align}
	\d f|_{A}(M)
		= & \sum_{j} \Pi_{j} M \Pi_{j}  \sum_{m=0}^{\infty} c_m m (\lambda_j - \alpha)^{m-1} \nonumber \\
		& + \sum_{j \neq k} \Pi_{j} M \Pi_{k} \sum_{m=0}^{\infty} c_m \frac{(\lambda_j-\alpha)^m - (\lambda_k-\alpha)^m}{\lambda_j - \lambda_k}.
\end{align}
Using the Taylor series expansions for $f(\lambda_j)$ and $f(\lambda_k)$, this simplifies to
\begin{equation}
	\d f|_{A}(M)
		= \sum_{j} f'(\lambda_{j}) \Pi_{j} M \Pi_{j}
			+ \sum_{j \neq k} \frac{f(\lambda_{j}) - f(\lambda_{k})}{\lambda_j - \lambda_k} \Pi_{j} M \Pi_{k}.
\end{equation}
This is equation \eqref{eq:locally-analytic-formula} from the main text.

\section{Tangent spaces to positive semidefinite operators} \label{app:psd}

Let $\mathcal{H}$ be an $n$-dimensional Hilbert space, and $\psd(\mathcal{H})$ the space of positive semidefinite operators.
A Hermitian operator $M$ is tangent to $\psd(\mathcal{H})$ at $\rho$ if the eigenvalues of $\rho + \epsilon M$ are all nonnegative at linear order in $\epsilon.$
For $M$ to be in the tangent \emph{space} of $\psd(\mathcal{H})$ at $\rho$, this same property must hold for all real multiples of $M$. 
The only time we can have $M$ tangent to $\psd(\mathcal{H})$ at $\rho$ but not in the tangent space of $\rho$ is when $\rho$ is on the boundary $\del \psd(\mathcal{H})$ -- if one of the zero-eigenvalues of $\rho$ becomes positive at linear order when perturbing by $M$, then $M$ is tangent to $\psd(\mathcal{H})$, but a perturbation by $-M$ will cause the zero-eigenvalue to become negative, so $-M$ is non-tangent to $\psd(\mathcal{H}).$

We conclude that for $\rho \in \psd(\mathcal{H})$, a Hermitian operator $M$ is in the tangent space $T_{\rho} \psd(\mathcal{H})$ if and only if the zero-eigenvalues of $\rho$ remain zero at linear order in the perturbation $\rho + \epsilon M.$ For general Hermitian $M$, the zero-eigenspace of $\rho$ will split into several eigenspaces at linear order in $\rho + \epsilon M.$ What this means is that the projector $\Pi_0$ onto the zero-eigenspace of $\rho$ can be decomposed into projectors
\begin{equation}
	\Pi_{0} = \sum_{j} \hat{\Pi}_{j},
\end{equation}
where each is continuously connected to an eigenspace projector of $\rho + \epsilon M$ by the formula
\begin{equation}
	\Pi_{j}^{(\rho + \epsilon M)} =\hat{\Pi}_{j} + \epsilon \delta \hat{\Pi}_{j}.
\end{equation}
The eigenvalue equation $(\rho + \epsilon M) \Pi_{j}^{(\rho + \epsilon M)} = \lambda_{j}^{(\rho + \epsilon M)} \Pi_{j}^{(\rho + \epsilon M)},$ evaluated at linear order in $\epsilon,$ gives
\begin{equation}
	\rho \delta \hat{\Pi}_{j} + M \hat{\Pi}_{j} = \delta \lambda_{j} \hat{\Pi}_{j}.
\end{equation}
Left-multiplying by $\hat{\Pi}_{k}$ and using $\hat{\Pi}_k \rho  = 0$ gives
\begin{equation}
	\hat{\Pi}_{k} M  \hat{\Pi}_{j} = \delta_{j k} \delta \lambda_{j} \hat{\Pi}_{j}.
\end{equation}
If $\mathcal{O}_{0}$ is a Hermitian operator that acts entirely within the zero-eigenspace of $\rho$, then we have
\begin{equation} \label{eq:eigenspace-orthogonality}
	\tr(M \mathcal{O}_{0}) =\tr(M \hat{\Pi}_0 \mathcal{O}_{0} \hat{\Pi}_0 ) = \sum_{j, k} \tr(M \hat{\Pi}_{j} \mathcal{O}_{0} \hat{\Pi}_{k}) = \sum_{j} \delta \lambda_{j} \tr( \hat{\Pi}_{j} \mathcal{O}_{0}).
\end{equation}
So we see that when $M$ is in $T_{\rho} \psd(\mathcal{H})$, meaning none of the zero-eigenvalues of $\rho$ change at linear order in $\rho + \epsilon M$, we must have $\delta \lambda_{j}=0$ and therefore $\langle M, \mathcal{O}_{0} \rangle_{\text{HS}} = 0$.

The argument of the preceding paragraph shows that whenever $M$ is tangent to $\psd(\mathcal{H})$ at $\rho$, it is orthogonal (in the sense of the Hilbert-Schmidt inner product) to every Hermitian operator acting on the zero-eigenspace of $\rho.$ Conversely, if $M$ is orthogonal to every operator in the zero-eigenspace of $\rho$, then it is in particular orthogonal to each projection operator $\hat{\Pi}_{j}$, which gives $\delta \lambda_{j} = 0$ for the corresponding eigenvalue by equation \eqref{eq:eigenspace-orthogonality}. We conclude that the tangent space to $\psd(\mathcal{H})$ at $\rho$ is exactly the set of Hermitian operators that are orthogonal to all operators acting on the zero-eigenspace of $\rho$. 

Another convenient way of conceptualizing this tangent space is that it is the space of all operators that can be written
\begin{equation} \label{eq:tangent-expression}
	M = \Delta \rho + \rho \Delta^{\dagger}
\end{equation}
for a general linear operator $\Delta \in \mathcal{L}(\mathcal{H}).$\footnote{We have not cited any sources in this appendix because the geometry of $T_{\rho} \psd(\mathcal{H})$ is farly simple and we did not consult any sources in deriving its properties; however, we credit chapter 5 of \cite{helmke2012optimization} for introducing us to the analogue of equation \eqref{eq:tangent-expression} for the space of positive semidefinite \emph{real} matrices.} It is straightforward to verify that every operator $M$ of the form given in equation \eqref{eq:tangent-expression} is orthogonal to every $\mathcal{O}_{0} \in \herm(E_0(\rho)),$ so every operator of the form \eqref{eq:tangent-expression} must lie within $T_{\rho} \psd(\mathcal{H}).$  The map
\begin{equation}
	\Delta \mapsto \Delta \rho + \rho \Delta^{\dagger}
\end{equation}
is an $\reals$-linear map from $\mathcal{L}(\mathcal{H})$ to $T_{\rho} \psd(\mathcal{H})$. A calculation we omit here shows that its kernel is $(n^2+k^2)$-dimensional, which tells us that its image must be $(2n^2 - (n^2+k^2))$-dimensional. This is the same dimension as $T_{\rho} \psd(\mathcal{H})$, so we conclude that \emph{every} $M$ in $T_{\rho} \psd(\mathcal{H})$ can be written in the form \eqref{eq:tangent-expression}.

\section*{References}
\bibliography{biblio}

\end{document}